\documentclass[journal]{IEEEtran}

\usepackage[utf8]{inputenc}
\usepackage{amsmath,amsfonts}
\usepackage{cite}
\usepackage{multicol}
\usepackage{graphicx}
\usepackage{array}
\usepackage{multirow}
\usepackage{graphicx}
\usepackage{enumitem}
\usepackage{color}
\usepackage{nomencl}
\usepackage{etoolbox}
\usepackage{amssymb}
\usepackage{comment}
\usepackage{mathtools}
\usepackage{bbold}
\usepackage{yfonts}
\usepackage{soul}
\usepackage{hyperref}
\usepackage{dsfont}
\usepackage[acronym]{glossaries}
\usepackage{nomencl}
\usepackage{subfiles}
\usepackage[utf8]{inputenc}
\usepackage{mathtools}
\usepackage{amsmath}
\usepackage{tikz}
\usepackage{algorithm}
\usepackage{algpseudocode}

\usepackage{fnpct}
\ifCLASSOPTIONcompsoc \usepackage[caption=false,font=normalsize,labelfon
t=sf,textfont=sf]{subfig}
\else
\usepackage[caption=false,font=footnotesize]{subfig}
\fi
\usepackage{tikz}

\title{Fairness-aware Photovoltaic Generation Limits for Voltage Regulation in Power Distribution Networks using Conservative Linear Approximations}
\author{
\IEEEauthorblockN{Rahul K. Gupta, Paprapee Buason, and Daniel K. Molzahn}\\
{School of Electrical and Computer Engineering, Georgia Institute of Technology,} Atlanta, USA \\
\{rgupta460, pbuason6, molzahn\}@gatech.edu
}

\usepackage{etoolbox}
\makeatletter
\patchcmd{\@maketitle}
  {\addvspace{0.5\baselineskip}\egroup}
  {\addvspace{-0.5\baselineskip}\egroup}
  {}
  {}
\makeatother
\begin{document}

\maketitle

\begin{abstract}
This paper proposes a framework for fairly curtailing photovoltaic (PV) plants in response to the over-voltage problem in PV-rich distribution networks. The framework imposes PV generation limits to avoid overvoltages. These limits are computed a day ahead of real-time operations by solving an offline stochastic optimization problem using forecasted scenarios for PV generation and load demand. The framework minimizes the overall curtailment while considering fairness by reducing disparities in curtailments among different PV owners. 
We model the distribution grid constraints using a conservative linear approximation (CLA) of the AC power flow equations which is computed using a set of sampled power injections from the day-ahead predicted scenarios. The proposed framework is numerically validated on a CIGRE benchmark network interfaced with a large number of PV plants. We compare the performance of the proposed framework versus an alternative formulation that does not incorporate fairness considerations. To this end, we assess tradeoffs between fairness, as quantified with the Jain Fairness Index (JFI), and the total curtailed energy. 
\end{abstract}
\begin{IEEEkeywords} 
Photovoltaic curtailment, Voltage regulation, Fairness-aware control, Conservative linear approximation.
\end{IEEEkeywords}

\vspace{-0.05em}
\section{Introduction}
Distribution system operators (DSOs) face the challenge of securely operating their grids amidst the rapid integration of photovoltaic (PV) plants. This involves ensuring the quality of supply (QoS) and adhering to the network's physical limits \cite{guide2004voltage, CIGREREF, IEEE_practice}. In the existing literature, this issue has been tackled by curtailment of the excess PV generation by real-time control schemes \cite{liu2020fairness, luthander2016self, von2018strategic, sevilla2018techno, o2020too}. However, such real-time controls require advanced communication and monitoring infrastructure which is not always available in distribution systems.
This issue has been often addressed by imposing fixed generation limits on the PV inverters to avoid over-voltage problems. For example, in \cite{aziz2017pv, ricciardi2018defining}, a percentage of the DC power module was used as the generation limit. 
In \cite{ricciardi2018defining}, the export limits were computed by formulating an optimal power flow (OPF) problem. However, in all the above works, the fairness aspect was not considered in the computation of these limits. 

Fairness is an important factor to consider while computing these generation limits since PV plants connected at different locations in the network face different grid conditions, leading to dissimilar curtailments. For example, customers located at the end of the feeder are likely to face more curtailments compared to ones near the substation. Therefore, in this paper, we propose to compute fairness-aware PV generation limits by incorporating a fairness objective in the optimization problem.

Recently, researchers have increasingly studied fairness in the context of control schemes. For example, different fairness-aware control schemes are proposed in \cite{liu2020fairness, gebbran2021fair, ali2015fair, gerdroodbari2021decentralized}. 
However, all these schemes rely on real-time control of the PV inverters which may not be practical in systems that lack sufficient communication infrastructure. In contrast, communicating PV generation limits once a day is more feasible since it does not require a low-latency communication network. 
Therefore, instead of real-time control, we propose computing offline PV generation limits (a day ahead of real-time operations) using a scenario-based stochastic optimization scheme that sends limits to the PV inverters once a day. To tackle disparities in curtailment among different PV owners, we add a fairness objective that minimizes differences in curtailments among consumers while respecting the grid constraints. We model the grid constraints using linearizations of the AC power flow equations known as ``conservative linear approximations'' (CLA) that are designed to respect grid constraints~\cite{BUASON2022}. The CLAs are computed using sampled active/reactive power injections, which in our case are obtained using day-ahead forecasts of the PV generation and load demands. Thanks to the CLAs, the optimization scheme is a linear program (LP) that can be effectively solved. We numerically validate the proposed scheme for a benchmark network connected with several PV plants. We also benchmark the proposed scheme with respect to a case that does not consider fairness. 

The key contributions of the proposed work are threefold. First, it develops an offline scenario-based stochastic optimization problem that can be solved and communicated once a day to curtail PV inverters. Second, it incorporates fairness in curtailments among different PV owners by minimizing disparities. Third, it employs the CLA approach to model grid constraints, linearizing the problem formulation while preserving the accuracy of the power flow model without causing violations.

The paper is organized as follows. 
Section~\ref{sec:problem formulation} describes the problem statement, grid model, PV plant model, and the day-ahead optimization problem. Section~\ref{sec:simulation} presents the numerical validation. Section~\ref{sec:conclusion} concludes the work.
\section{Problem Formulation} \label{sec:problem formulation}
\begin{figure}[!htbp]
    \centering
    \includegraphics[width=\linewidth]{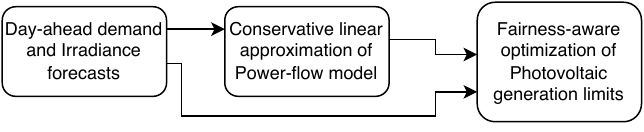}
    \caption{Flow chart for the computation of day-ahead PV generation limits.}
    \label{fig:flow}
\end{figure}

We consider a generic distribution network interfaced with multiple PV plants facing the problem of over-voltages in the case of substantial PV generation. The problem is tackled via curtailing the PV generation by imposing limits on the PV inverters. To avoid real-time communication and monitoring requirements, we propose offline computations of PV generation limits which are performed a day ahead of real-time operation based on forecasted scenarios. The proposed scheme accounts for fairness among different PV owners with respect to the curtailment by adding a fairness objective in the optimization problem. To model the grid constraints, we use conservative linear approximations (CLA) of the power flow equations, resulting in a linear problem formulation. The proposed scheme is schematically shown in Fig.~\ref{fig:flow}; first the day-ahead scenarios are used to calculate the CLA coefficients using the method described in \cite{BUASON2022}, and then the scenarios are used to compute the PV generation limits.

In the next subsections, we describe the models used for the distribution grid, PV generation, and demand. Then, we present our proposed fairness-aware day-ahead optimization problem for computing the PV generation limits.

\subsection{Grid model} 
\label{sec:grid}
We employ linearizations of the power flow equations known as \emph{conservative linear approximations} (CLAs) to address challenges arising from the nonlinearity of power flow constraints. These CLAs, introduced in~\cite{BUASON2022}, are adaptive (i.e., tailored to specific systems and varying load conditions) and conservative (i.e., aim to either over- or under-estimate a quantity of interest to prevent constraint violations). The CLAs are constructed using samples of the active and reactive power injections over a specified range of operations.
As described in \cite{BUASON2022}, the CLAs' coefficients are computed by solving a constrained regression problem, establishing linear relationships between, for example, the voltage magnitudes\footnote{The CLA method can be applied to compute approximations for other quantities of interest, including line flows, using the same approach. This paper uses CLAs to model voltages.} at a particular node and the power injections at all nodes.

Rather than solely considering the voltage magnitudes themselves, the CLA approach's accuracy can be enhanced by considering \emph{functions} of voltage magnitudes, such as squared voltage magnitudes ($v^2$). These approximations maintain linearity in constraints (e.g., $v \leq v_\text{max}$ implies $v^2 \leq v_\text{max}^2$); see \cite{BUASON2022} for further details. Since approximations of the squared voltage magnitudes tend to perform better than approximations of the voltage magnitudes themselves, we use these for our constraints. 
For the $j$--th node in a distribution network with $N_b$ nodes, the constraints for over- and under-estimated squared voltages using CLA are
\begin{subequations}
\begin{align}
    & \bar{v}_j^2 = \bar{a}_{j,0} + \bar{\mathbf{a}}_{j,1}^\top
    \begin{pmatrix}
        \mathbf{p} \\  
        \mathbf{q}
    \end{pmatrix} \leq v_\text{max}^2, \label{eq:CLA_overestimate}\\
    & \underline{v}_j^2 = \underline{a}_{j,0} + \underline{\mathbf{a}}_{j,1}^\top
    \begin{pmatrix}
        \mathbf{p} \\  
        \mathbf{q}
    \end{pmatrix} \geq v_\text{min}^2 \label{eq:CLA_underestimate},  
\end{align}%
\label{eq:CLA}%
\end{subequations}%
where $\bar{v}_j^2$ and $\underline{v}_j^2$ denote over- and under-estimating CLAs parameterized with coefficients $\bar{a}_{j,0} \in \mathbb{R}, \bar{\mathbf{a}}_{j,1} \in \mathbb{R}^{2(N_b-1)}$, and $\underline{a}_{j,0} \in \mathbb{R}, \underline{\mathbf{a}}_{j,1} \in \mathbb{R}^{2(N_b-1)}$, respectively. The symbols $v_\text{min}$ and $v_\text{max}$ represent the minimum and maximum voltage bounds. The superscript $\top$ denotes the transpose. The symbols $\mathbf{p} \in \mathbb{R}^{(N_b-1)}$ and $\mathbf{q} \in \mathbb{R}^{(N_b-1)}$ denote the nodal active and reactive power injection vectors (excluding the slack node injections).
Equations~\eqref{eq:CLA_overestimate} and \eqref{eq:CLA_underestimate} imply that if the overestimating and underestimating voltages are within the limits, the actual voltage remains within the limits, assuming the computed CLAs are indeed conservative.

In our example involving voltages, we compute sample-based CLAs by randomly sampling power generation outputs from PV plants and demand within a specified range. These samples are derived from the day-ahead predictions of the PV generation and loads. Using these samples, we calculate the corresponding node voltages by solving the power flow equations for each sample. The power generation samples in day-ahead PV scenarios are obtained using specified probability distributions. 
The active power injections are sampled from a uniform probability distribution function with active power varying between 0 and $\hat{p}^\text{pv}_{\text{max}}$, where $\hat{p}^\text{pv}_{\text{max}}$ is maximum PV power available (subject to solar irradiance). The reactive power from the PV plant is modeled as 33\% of the active power potential (corresponding to the imposed power-factor of 0.95). Section~\ref{sec:pf_validation} discusses the CLA's accuracy.

\subsection{PV model}
\label{sec:PVmodel}
In the problem formulation, we model PV generation using day-ahead forecasts derived from the global horizontal irradiance (GHI) and air temperature predictions that are available from standard weather forecasting tools, e.g., Solcast \cite{bright2019solcast}. To convert the GHI to the PV generation, we utilize already existing PV-lib models~\cite{holmgren2018pvlib, sossan2019solar}. This involves projecting GHI onto an inclined plane using the PV panels' tilts (assumed to be known in this work). Then, the inclined GHI and temperature data is fed to the PV-lib model which also requires the configuration of the PV plant such as nominal power and azimuth angles (known parameters in the forecasting model). The obtained PV generation forecasts, denoted by $\hat{{p}}^\text{pv}$, provide the maximum power point (MPP) for a PV plant given the GHI. Let the uncertain scenarios for a PV plant be indexed by $\omega \in \Omega$, where $\Omega$ represents the scenario set. Let $t \in \mathcal{T}$ represent the time index and $\mathcal{T}$ be the set of time indices during a day. The PV plants are indexed by symbol $j$ contained in set $\mathcal{N}_{\text{pv}}$. Then, the MPP of the $j$--th PV plant for time $t$ and scenario $\omega$ is denoted by $\hat{p}_{j,t, \omega}^\text{pv}$.

For the case of voltage regulation, we consider PV plants to be controllable, i.e., their active power generation can be reduced from the available peak power (i.e., MPP). We also consider controllable reactive power injections. Let the symbols $p_{j,t, \omega}^{\text{pv}}$ and $q_{j,t, \omega}^{\text{pv}}$ denote the active and reactive powers for the $j$--th PV plant at time $t$ in scenario $\omega$. 
Then, the curtailability is defined by the following constraint which states that the generation can vary between 0~kW and the generation potential:
\begin{subequations}
\label{eq:PV_model}
\begin{align}
    {0} \leq {p}_{j, t, \omega}^{\text{pv}}  \leq \widehat{{p}}_{j, t, \omega}^{\text{pv}} && \forall t, \omega, j.\label{eq:pv_activeP}
\end{align}

The PV power plants can also inject/consume reactive power, and we thus have an additional constraint on the capability curve of the PV plant's inverter that limits the apparent power capacity to $S^\text{pv}_{j,\text{max}}$:
\begin{align*}
    & ({p}_{j, t, \omega}^{\text{pv}})^2 + ({q}_{j, t, \omega}^{\text{pv}})^2 \leq  ({S}^{\text{pv}}_{j,\text{max}})^2 && \forall t, \omega, j. 
\end{align*}

If a linear formulation is desired, the quadratic capability constraint can be piece-wise linearized by re-writing it as:
\begin{align}
    & m_l ({p}_{j, t, \omega}^{\text{pv}}) + {q}_{j, t, \omega}^{\text{pv}} \leq  n_l && l =1, \dots, L, &  \forall t, \omega, j,  \label{eq:capability_linear}
\end{align}
where $m_l$ and $n_l$ denote the linearization coefficients for $l$--th segment in the piece-wise linearization and $L$ denotes the number of linear segments.

Furthermore, we also consider the minimum power factor constraint (for simplicity, we assume that all the PV plants have the same power factor):
\begin{align}
    &  {q}_{j, t, \omega}^{\text{pv}} \leq {p}_{j, t, \omega}^{\text{pv}}\zeta && \forall t, \omega, j \label{eq:pf1_ch6}\\
    &  -{q}_{j, t, \omega}^{\text{pv}}  \leq {p}_{j, t, \omega}^{\text{pv}}\zeta && \forall t, \omega, j \label{eq:pf2_ch6},
\end{align}
where $\zeta = \sqrt{(1-\text{PF}^2_{\text{min}})/\text{PF}^2_{\text{min}}}$ with $\text{PF}_{\text{min}}$ denoting the minimum power factor allowed for the operation of each PV plant. For the sake of simplicity, we consider the same minimum power factor of 0.95 on all PV plants in the network.
\end{subequations}

\subsection{Demand model}
\label{sec:demand}
We model the electricity demand to be uncontrollable but predictable based on day-ahead forecasts. We use the multivariate Gaussian model proposed in \cite{gupta2020grid, gupta2022reliable} for demand forecasting. This model is constructed using historical measurements that are clustered into different day types based on the day of the week. For each cluster, a multivariate distribution is trained considering time correlations, which is then used for sampling new scenarios. The scenarios for active and reactive loads for node index $n \in \mathcal{N}_\text{load}$ are denoted by $p^\text{load}_{n, t, \omega}$ and $q^\text{load}_{n, t, \omega}$, respectively for time $t$ and scenario $\omega$.

\subsection{Day-ahead optimization problem for fairness-aware PV generation limits}
We consider the distribution network to have $N_{\text{pv}}$ controllable PV plants indexed by $j$ in set $\mathcal{N}_{\text{pv}} \subset \mathcal{N}_b$, where $\mathcal{N}_b = \{1, \dots, N_b\}$ is the set of all node indices. The objective of the day-ahead optimization problem is to compute the PV generation limits $\bar{p}^{\text{pv}}_j$ for each PV plant to mitigate overvoltages. These limits should account for the uncertainty in the PV generation and electricity demand and also be fair with respect to the PV curtailment among different PV owners. 
The problem is formulated as scenario-based stochastic optimization where the scenarios model uncertainties in both demand and generation. 
Let the vectors $\mathbf{p}_{j, \omega}^{\text{pv}} \in \mathbb{R}^{|\mathcal{T}|}$ and $\widehat{\mathbf{p}}_{j, \omega}^{\text{pv}} \in \mathbb{R}^{|\mathcal{T}|}$ be the decision variables and MPP, respectively across all timesteps for the $j$--th PV plant in scenario $\omega$. The PV active power generation is constrained by the PV generation limit $\bar{p}^{\text{pv}}_j$, which is also a decision variable. The constraint is
\begin{align}
    p_{j,t,\omega}^\text{pv} \leq \bar{p}^{\text{pv}}_j, && \forall t,\omega,j.%
    \label{eq:PVlimit}%
\end{align}%
The optimization problem minimizes a multi-objective function given by\footnote{$\|.\|_p$ refers to p-norm.}
\begin{align}
\begin{aligned}
      f^\text{op}(\mathbf{x}, {\Theta}) = \sum_{j\in\mathcal{N}_{\text{pv}}} & \Bigg\{\alpha_1 \times  \Big( \sum_{\omega \in \Omega} \big\|\mathbf{p}_{j, \omega}^{\text{pv}}-\widehat{\mathbf{p}}_{j, \omega}^{\text{pv}}\big\|_1+ \textcolor{black}{\bar{p}^{\text{pv}}_j} \Big)  + \\ & + \alpha_2 \times \sum_{\omega \in \Omega}\Big\|\gamma_{\omega} - \Gamma(\mathbf{p}_{j, \omega}^{\text{pv}}, \widehat{\mathbf{p}}_{j, \omega}^{\text{pv}}) 
     \Big\|_2 \Bigg\}.
\end{aligned}
\label{eq:obj}
\end{align} 
The symbol $\mathbf{x}$ collects all the decision variables, i.e., $\mathbf{x} = [ \mathbf{p}_{j, \omega}^{\text{pv}}$, $\forall j, \omega$, \textcolor{black}{$\bar{p}^{\text{pv}}_j, \forall j$]}, and $\Theta$ refers to the set of parameters, i.e, $\alpha_1, \alpha_2$ and $\widehat{\mathbf{p}}_{j, \omega}^{\text{pv}}$, $\forall j, \omega$.
In \eqref{eq:obj}, the first term (weighted by $\alpha_1$) minimizes the curtailments with respect to the available MPP (given by the day-ahead forecasts $\hat{\mathbf{p}}_{j,\omega}^{\text{pv}}$). 
\textcolor{black}{The term $\bar{p}^{\text{pv}}_j$ in the objective ensures that the constraint in \eqref{eq:PVlimit} is binding.}
The second term (weighted by $\alpha_2$) in \eqref{eq:obj} seeks to impose fairness in curtailments across PV plants. Here, $\gamma_{\omega}$ is a common metric (per scenario) that enforces fairness among different PV plants using a fairness function $\Gamma$. The function $\Gamma$ can be any generic metric to enforce fairness among different PV plant owners; in this work, we define $\Gamma$ as the ratio of the total energy produced and maximum available energy, i.e., 
\begin{align}
    \Gamma(\mathbf{p}_{j,\omega}^{\text{pv}}, \widehat{\mathbf{p}}_{j, \omega}^{\text{pv}}) = \frac{\sum_{t\in \mathcal{T}}{p_{j,t, \omega}^{\text{pv}}}}{\sum_{t\in \mathcal{T}}{\widehat{p}_{j,t, \omega}^{\text{pv}}}}.
\end{align}
The symbols $\alpha_1$ and $\alpha_2$ are weights\footnote{A sensitivity analysis with different values of the weights is presented by the results in Section~\ref{sec:sensitivity}.} corresponding to the two objectives. 
The day-ahead optimization problem is 
\begin{subequations}
\label{eq:dayahead_OP}
\begin{align}
      &\underset{\mathbf{x}}
      {\text{minimize}}~ f^{op}(\mathbf{x}, {\Theta})
      \end{align}
subject to:
{\color{black}
      \begin{align}
    & \bar{a}_{j,0} + \bar{\mathbf{a}}_{j,1}^\top
    \begin{pmatrix}
        \mathbf{p_{t,\omega}} \\  
        \mathbf{q}_{t, \omega}
    \end{pmatrix} \leq v_\text{max}^2, && \forall t, \omega, j\\
    & \underline{a}_{j,0} + \underline{\mathbf{a}}_{j,1}^\top
    \begin{pmatrix} 
        \mathbf{p}_{t,\omega} \\  
        \mathbf{q}_{t,\omega}
    \end{pmatrix} \geq v_\text{min}^2,  && \forall t, \omega, j\\
    & \eqref{eq:PV_model}, \, \eqref{eq:PVlimit},
\end{align}}
\end{subequations}
\textcolor{black}{where the symbols $\mathbf{p}_{t,\omega} \in \mathbb{R}^{(N_b-1)}$ and $\mathbf{q}_{t,\omega} \in \mathbb{R}^{(N_b-1)}$ denote the nodal active and reactive power injection vectors for time index $t$ and scenario $\omega$. These nodal vectors contain the power injections for $N_\text{pv}$ controllable PV plants and $N_\text{load}$ uncontrollable loads as per their locations defined by sets $\mathcal{N}_{\text{pv}}$ and $\mathcal{N}_{\text{load}}$, respectively.}

\begin{figure*}[!t]
    \centering
    \includegraphics[width=0.71\linewidth]{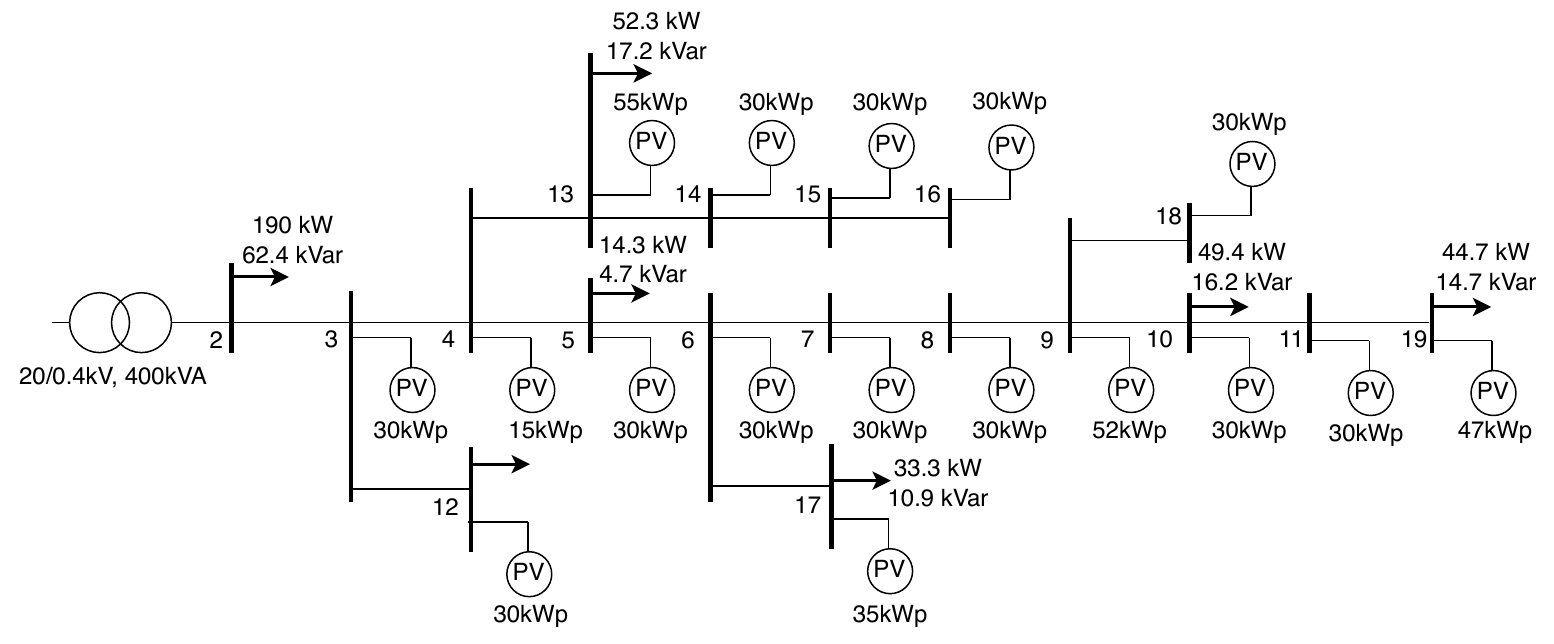}
    \caption{CIGRE benchmark low-voltage distribution grid with multiple PV plants that result in over-voltage problems.}
    \label{fig:griddata}
\end{figure*}

\section{Numerical Validation}
\label{sec:simulation}
This section numerically validates the proposed framework. We describe the distribution system test case, the scenarios used for the simulations, and the obtained results.
\subsection{Grid data}
We numerically validate the proposed framework on the CIGRE low-voltage benchmark distribution network \cite{CIGREREF} augmented with multiple PV plants so that the system experiences over-voltage problems. The network is a three-phase balanced 400V/400kVA system. The network also hosts uncontrollable active and reactive demands. Fig.~\ref{fig:griddata} illustrates the locations and nominal capacities of PV plants and demands.
\subsection{Day-ahead scenarios}
As mentioned previously, our framework relies on day-ahead scenarios for PV generation and demand. While the optimization problem is generic enough to consider any number of scenarios, we represent the uncertainty of the generation and demand by two extreme scenarios.
\begin{figure}[!b]
\centering
\subfloat[PV generation potential (MPP) scenarios for different nodes.]{\includegraphics[width=\linewidth]{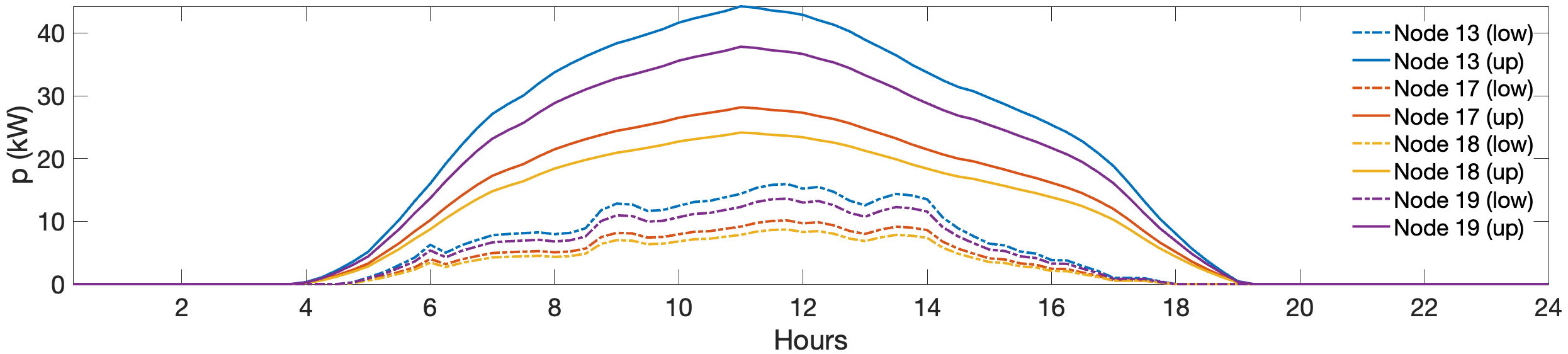}
\label{fig:PV}}\\
\subfloat[Active demand scenarios for different nodes.]{\includegraphics[width=\linewidth]{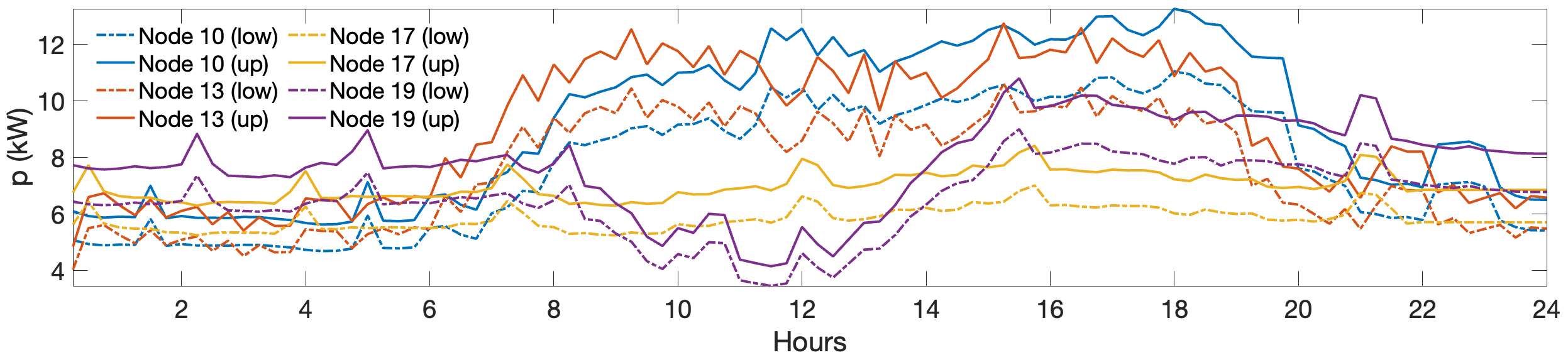}
\label{fig:loadP}}\\
\subfloat[Reactive demand scenarios for different nodes.]{\includegraphics[width=\linewidth]{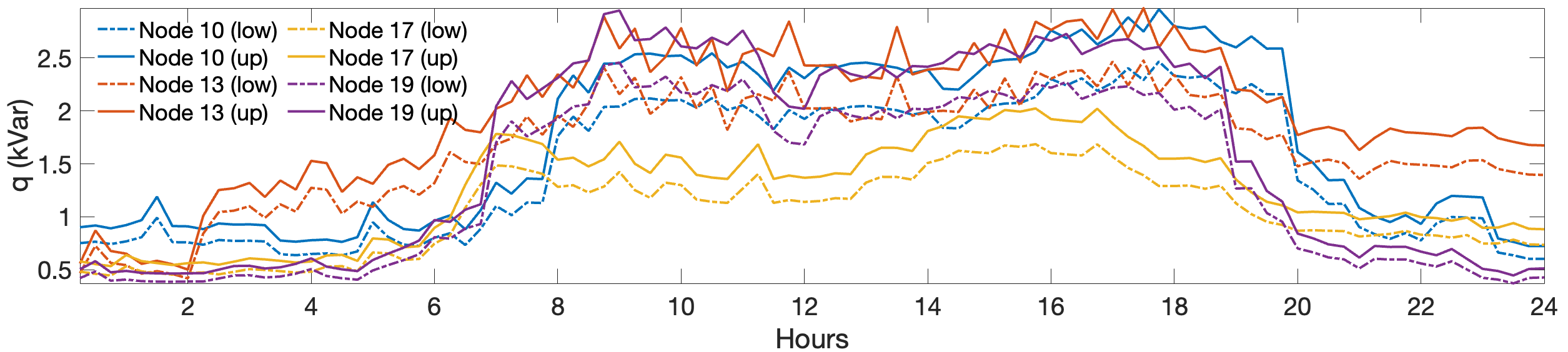}
\label{fig:loadQ}}\\
\caption{Day-ahead scenarios of (a) PV generation, (b) active power demand, and (c) reactive power demand. The tags `low' and `up' correspond to the 10-th and 90-th percentile of the predictions in each case.} \label{fig:dayahead_sc}
\end{figure}
The PV generation is computed using the day-ahead GHI forecasts represented by their upper and lower predictions corresponding to the 90-th and 10-th percentile. The GHI profiles are obtained from Solcast \cite{bright2019solcast} at a sampling of 15-minutes. These GHI profiles are then used to compute the MPP of each PV plant using the PV model described in Section~\ref{sec:PVmodel}. The computed MPP profiles for the upper and lower scenarios for a subset
of the nodes are shown in Fig.~\ref{fig:PV}.

We also model the demand by two extreme profiles corresponding to the 90-th and 10-th percentiles. These scenarios are obtained at a sampling of 15-minutes using the forecasting approach described in Section~\ref{sec:demand}, and are shown in Figs.~\ref{fig:loadP} and \ref{fig:loadQ} for the active and reactive power demands, respectively. To consider the extreme case with respect to the voltage problem in the distribution network, we couple the low PV generation scenario with high demand and vice-versa, and then we choose the two resulting extreme scenarios.

Note that these day-ahead scenarios of the PV generation and demand are used as inputs for sampling the power injections per node when computing the CLA coefficients as described in Section~\ref{sec:grid}.

\subsection{Results}
We next present the results simulated for the CIGRE benchmark network. We compare the performance against the case when fairness is not considered in the objective. Then, we perform sensitivity analyses with respect to the choice of the weights ($\alpha_1,\alpha_2$) in terms of fairness and curtailments. Finally, we validate the CLA approach against the true non-linear AC power flow solutions. The results are \emph{only} presented for the over-voltage case, i.e., for the day-ahead scenario corresponding to larger PV generation and lower demand. \textcolor{black}{We do not present the results of the other day-ahead scenario, as it did not cause any voltage issues, i.e., no PV curtailments were required.}
\subsubsection{Without and with fairness}
\begin{figure}[t]
\centering
\subfloat[Curtailed and MPP PV generation.]{\includegraphics[width=\linewidth]{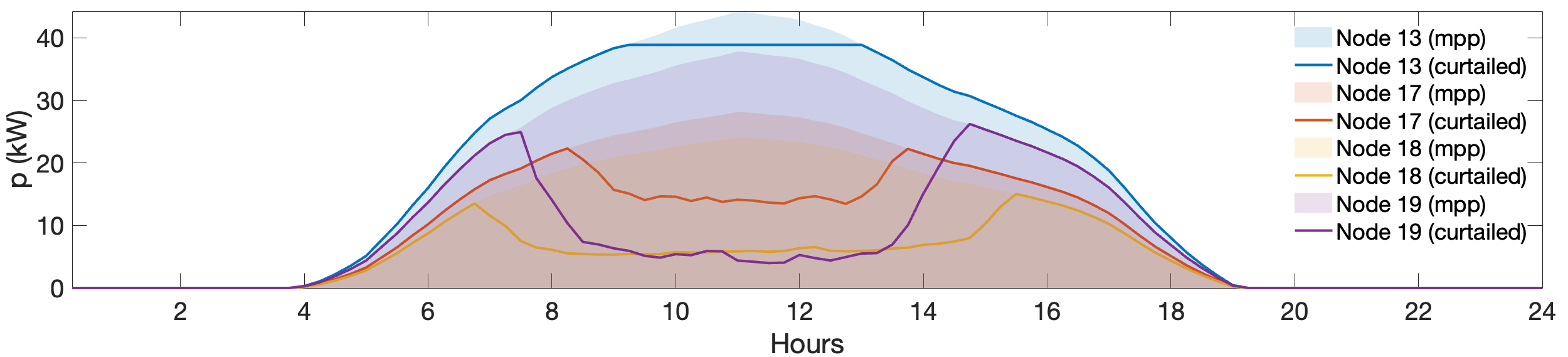}
\label{fig:PVp_unfair}}\\
\subfloat[Reactive power from PV plants.]{\includegraphics[width=\linewidth]{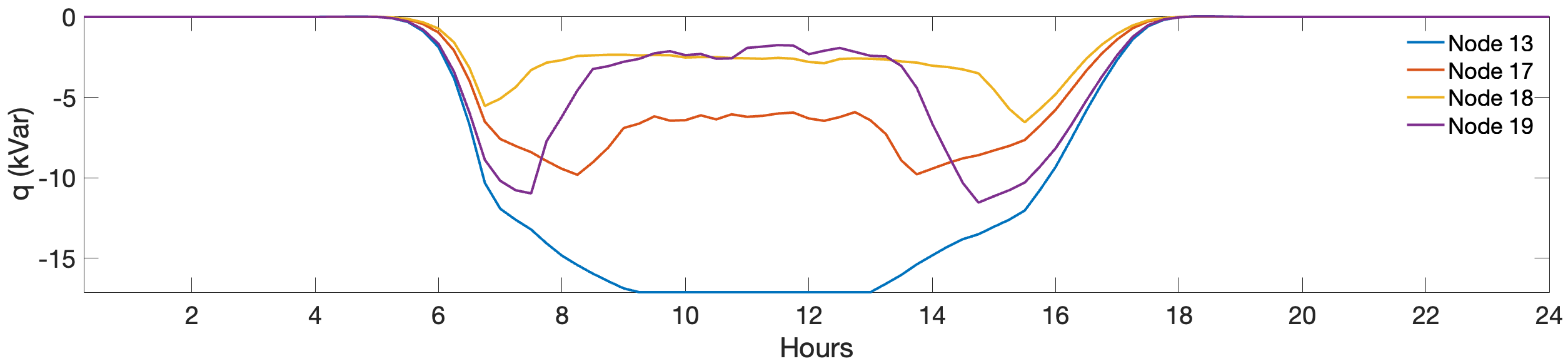}
\label{fig:PVq_unfair}}\\
\subfloat[Nodal voltage magnitudes.]{\includegraphics[width=\linewidth]{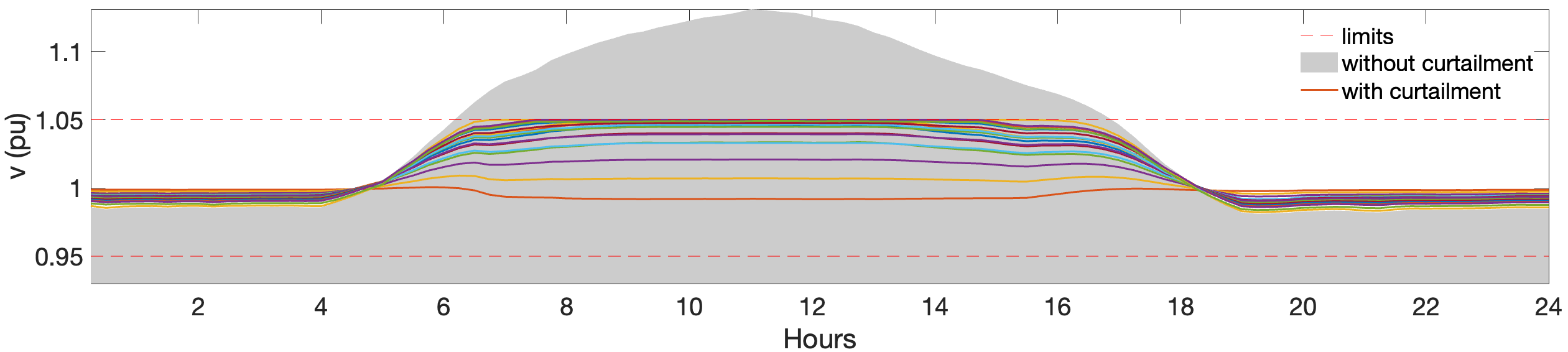}
\label{fig:V_unfair}}\\
\caption{Simulation results for the case when fairness is not considered i.e., ($\alpha_2 = 0$).} 
\label{fig:unfair}
\end{figure}

\begin{figure}[t]
\centering
\subfloat[Curtailed and MPP PV generation.]{\includegraphics[width=\linewidth]{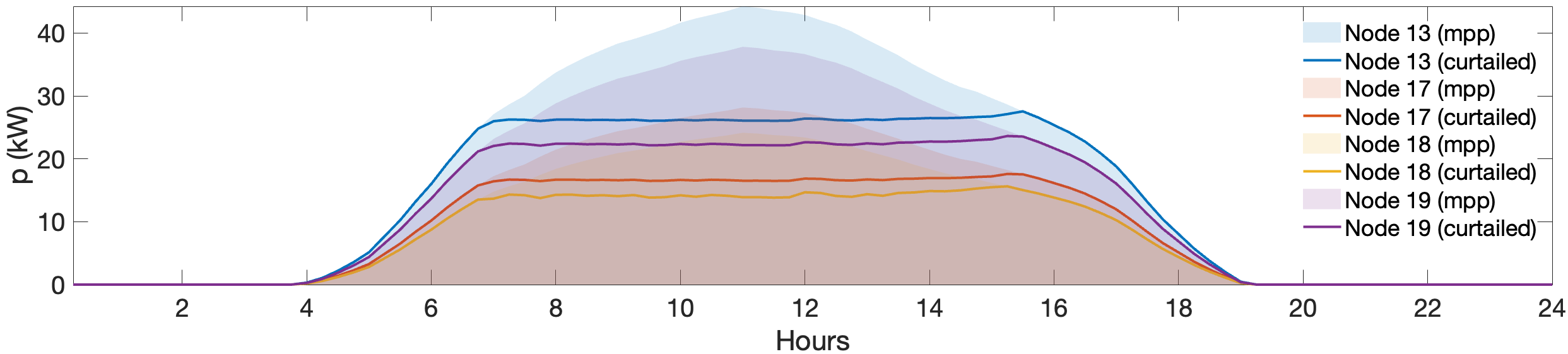}
\label{fig:PVp_fair}}\\
\subfloat[Reactive power from PV plants.]{\includegraphics[width=\linewidth]{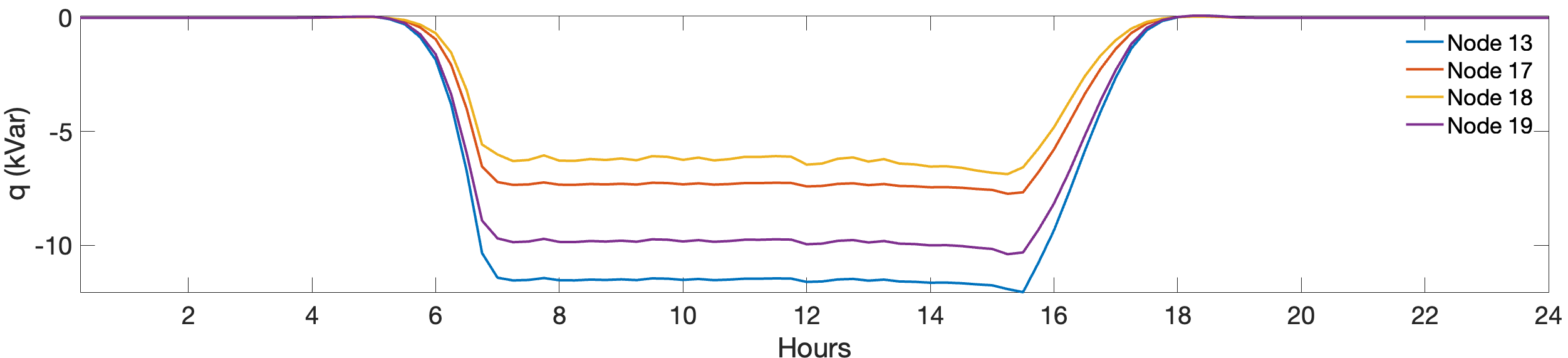}
\label{fig:PVq_fair}}\\
\subfloat[Nodal voltage magnitudes.]{\includegraphics[width=\linewidth]{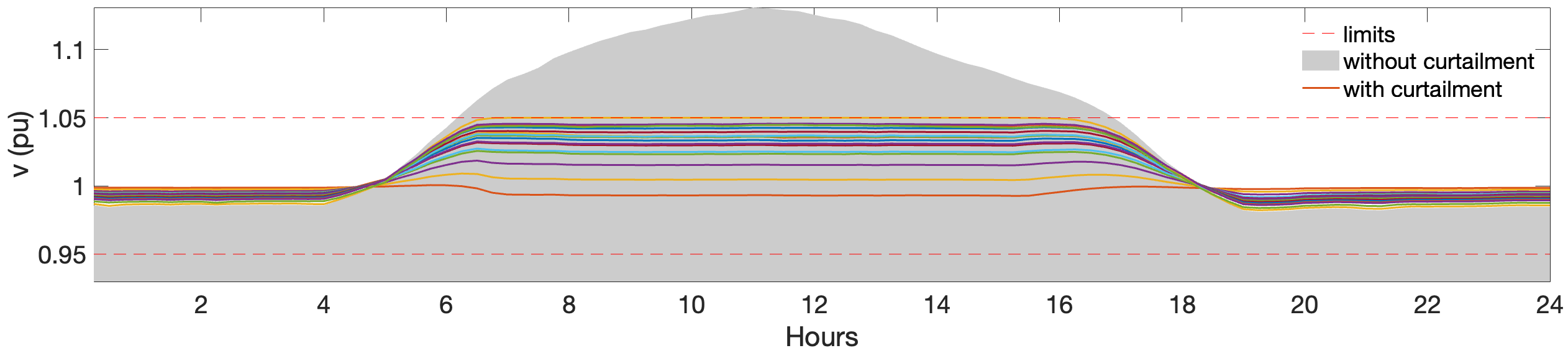}
\label{fig:V_fair}}\\
\caption{Simulation results for the case with fairness ($\alpha_2 = 7$).} 
\label{fig:fair}
\end{figure}

Figs.~\ref{fig:unfair} and \ref{fig:fair} present the results when the fairness objective is not considered (i.e., $\alpha_2 = 0$) and when fairness is considered ($\alpha_2 = 7$), respectively. The first panels (Figs.~\ref{fig:PVp_unfair} and \ref{fig:PVp_fair}) show the curtailed PV and corresponding MPP in lineplots and shaded areas, respectively. The second panels (Figs.~\ref{fig:PVq_unfair} and \ref{fig:PVq_fair}) show the reactive power injections by the PV inverters. The bottom panels (Figs.~\ref{fig:V_unfair} and \ref{fig:V_fair}) show the nodal voltage magnitudes. Lineplots (shaded grey) correspond to voltages with (without) PV curtailment.

\begin{figure}[!t]
    \centering
    \includegraphics[width=\linewidth]{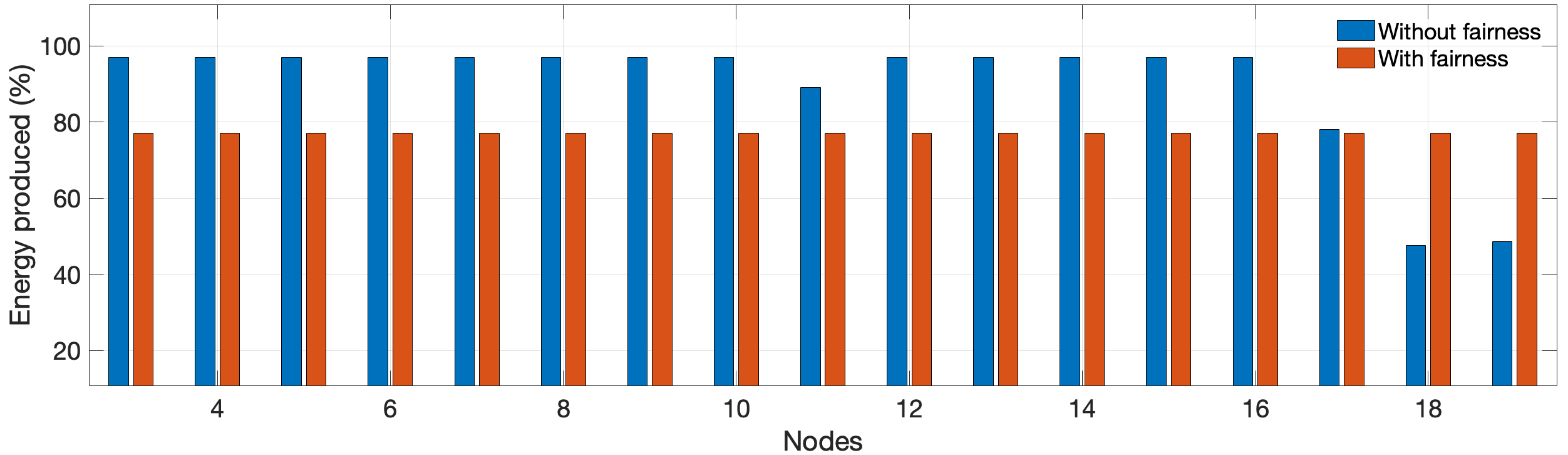}
    \vspace{-2em}
    \caption{Comparison of energy produced by PV plants with and without considering fairness.}
    \label{fig:PVE_compare}
\end{figure}

As seen from the Figs.~\ref{fig:V_unfair} and~\ref{fig:V_fair}, the PV curtailment results in satisfying the voltage limits of [0.95, 1.05] per unit in both cases. However, the PV active and reactive profiles in the two cases are different. By comparing the plots in Figs.~\ref{fig:PVq_unfair} and \ref{fig:PVq_fair}, we observe that some PV plants suffer significantly higher curtailment (in the first case) compared to others, whereas in the case of fairness-aware curtailment, all the PV plants are equally curtailed. This is evident from the comparison of the energy produced as a percentage of the maximum available energy, which is shown via the barplot in Fig.~\ref{fig:PVE_compare}. Specifically, without considering fairness concerns, the PV plants located at nodes 11, 17, 18, and 19 suffer significantly higher curtailment than others. In contrast, all of the PV plants are equally curtailed in the fairness-aware case.

To quantify fairness, we use Jain's Fairness Index (JFI) \cite{jain1984quantitative} which is a metric to quantify the spread of benefits to each consumer using different control schemes. JFI values vary between 0 and 1, where JFI = 0 and JFI = 1 refer to completely unfair and fair cases, respectively.
The JFI is given by  
\begin{align}
    \text{JFI} = \frac{(\sum_{j\in\mathcal{N}_\text{pv}}\beta_j)^2}{|\mathcal{N}_{\text{pv}}|{\sum_{j\in\mathcal{N}_\text{pv}}\beta_j^2}}
\end{align}
where $\beta_j$ refers to the percentage of PV energy produced, i.e., 
\begin{align}\label{eq:beta_def}
    \beta_j = \Gamma(\mathbf{p}_{j,\omega}^{*\text{pv}}, \widehat{\mathbf{p}}_{j, \omega}^{\text{pv}}) = \frac{\sum_t{p_{j,t, \omega}^{*\text{pv}}}}{\sum_t{\widehat{p}_{j,t,\omega}^{\text{pv}}}}.
\end{align}
In~\eqref{eq:beta_def}, the asterisk symbol $(\,\cdot\,)^*$ refers to the solution to optimization problem~\eqref{eq:dayahead_OP}. 

Using the above fairness index, Table~\ref{tab:comaprison} compares the JFI and net curtailment (total curtailed PV as a percentage of total available energy) for the two cases. 
\begin{table}[!b]
    \centering
    \caption{Net PV Curtailments with and without fairness.}
    \begin{tabular}{|c|c|c|}
    \hline 
    Metrics             & Net curtailment (\%) & JFI\\
    \hline
    \hline
    Without Fairness    &     11.25          &       0.29    \\
    With Fairness      &       23.01         &       1    \\
    \hline
    \end{tabular}
    \label{tab:comaprison}
\end{table}
Although the fairness-aware scheme eliminates the disparity in curtailment among different PV owners, the result is larger  overall net curtailments. Therefore, in the following section, we present a sensitivity analysis with respect to the choice of $\alpha_2$ to evaluate the trade-offs between curtailment and fairness.
\begin{figure}[!b]
    \centering
    \includegraphics[width=0.6\linewidth]{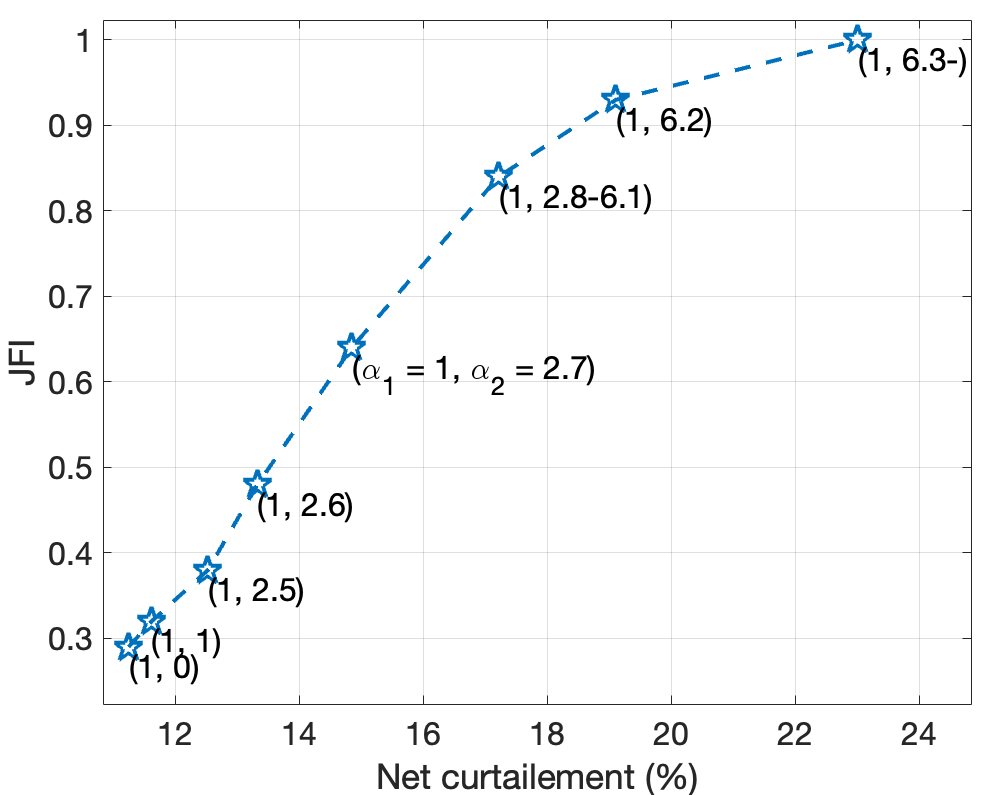}
    \caption{Sensitivity of fairness vs. net curtailment with respect to weight $\alpha_2$.}
    \label{fig:sensitivity}
\end{figure}
\begin{figure}[!htbp]
    \centering
    \includegraphics[width=0.63\linewidth]{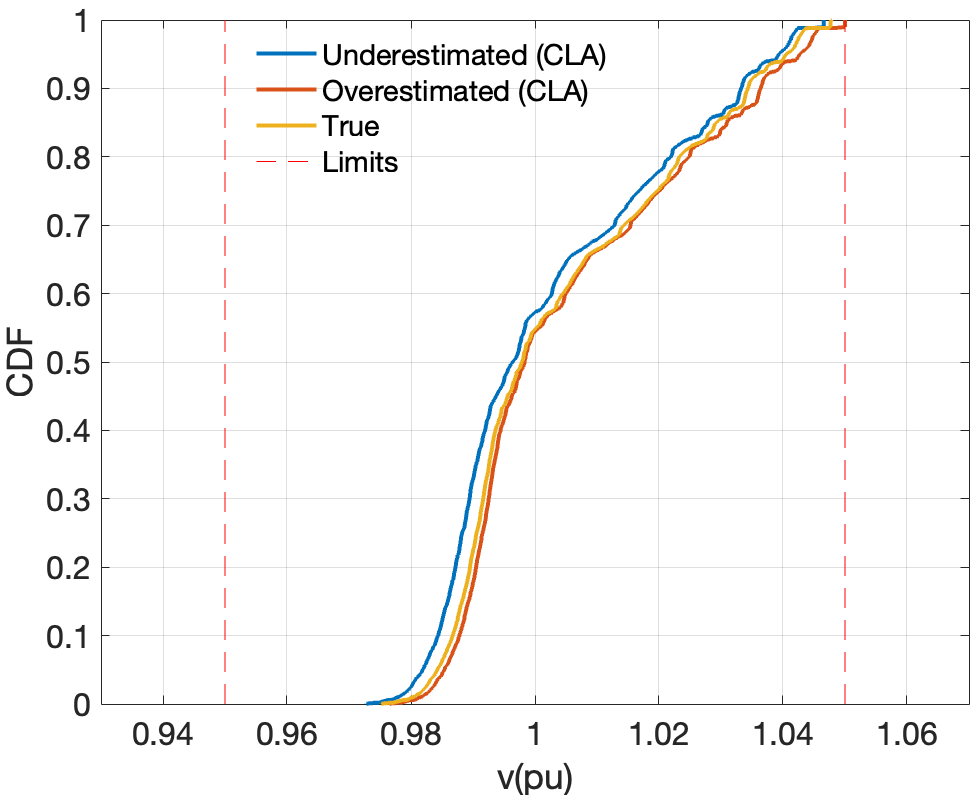}
    \caption{Validation of CLA approximation of the voltage model with respect to true values (obtained by solving the AC power flow equations, post-optimization, using the optimized power setpoints).}
    \label{fig:CLA}
\end{figure}
\subsubsection{Fairness-curtailment trade-off: Sensitivity with respect to the weights}
\label{sec:sensitivity}
We compute the JFI and net curtailed energy while varying the weighing factor $\alpha_2$ from 0 (i.e., no fairness) to 10. The obtained results are shown in Fig.~\ref{fig:sensitivity}. As expected, JFI increases with an increase in $\alpha_2$, at the cost of increasing net PV curtailment. The JFI value saturates at 1.0 (i.e., full fairness) after $\alpha_2 = 6.3$.

\subsubsection{Power flow accuracy validation}
\label{sec:pf_validation}
We also validate the CLA approximation of the power flow equation which is used to model the grid constraints in the day-ahead optimization problem in \eqref{eq:dayahead_OP}. The validation is performed by comparing the optimized solutions with respective ``true'' solutions obtained by solving the AC power flow using the optimized solution.

Fig.~\ref{fig:CLA} shows the comparison of under- and over-estimated voltages by CLAs ($\underline{v}$ and $\bar{v}$) and the true values (denoted by $v^\text{true})$. As observed, all the voltages are within the imposed limits of 1.05, and voltages computed by CLAs are close to the true values. \textcolor{black}{This indicates that there are no voltage violations, concluding that linear approximations by CLAs perform as intended.} Table~\ref{tab:CLA_errors} shows the minimum, mean, and maximum errors on the nodal voltage magnitudes. Moreover, the errors in the voltage magnitudes computed by the CLA are quite small, indicating that CLAs provide adequate accuracy for the presented optimization framework.
\begin{table}[!htbp]
    \centering
    \caption{Errors on voltage magnitudes computed by CLA.}
    \begin{tabular}{|c|c|c|c|}
    \hline 
      CLA errors   & Min (pu) & Mean (pu) & Max (pu) \\
    \hline
    \hline
     $\bar{v} - v^{\text{true}}$ \emph{(over-estimated)}   &  $1\times 10^{-4}$   &  $1\times 10^{-3}$    &   $2.4\times 10^{-3}$   \\
     $v^{\text{true}} - \underline{v} $ \emph{(under-estimated)}  &  $4\times 10^{-3}$   &   $2\times 10^{-3}$   &   $3.2\times 10^{-3}$   \\
    \hline
    \end{tabular}
    \label{tab:CLA_errors}
\end{table}
\vspace{-1em}
\section{Conclusions}
\label{sec:conclusion}
This paper proposed an offline optimization framework for computing fairness-aware PV generation limits for PV inverters in power distribution networks to address over-voltage problems. Fairness is accounted for by adding an extra term in the objective which minimizes the disparity in the amount of curtailment among different PV plants generation. The problem is solved a day ahead of real-time operation via a scenario-based stochastic optimization problem where the scenarios model uncertainties in PV generation and load demands. The grid constraints are accounted for by a conservative linear approximation of the AC power flow equations, making the problem a linear-constrained program.

The proposed scheme was numerically validated on a CIGRE low-voltage network. The obtained results showed that the proposed framework successfully achieves fairness with respect to the individual curtailment among different PV owners while respecting the imposed voltage limit. However, it was also observed that increasing fairness leads to increased overall curtailment in the PV generation. Future work will focus on developing strategies for improving the trade-offs between the overall net curtailment and fairness considerations.

\bibliographystyle{IEEEtran}
\bibliography{bibliography.bib}
\end{document}